\documentclass[prl,aps,twocolumn]{revtex4}
\usepackage{graphicx}
\begin{document}
\title{Observation of angle-dependent transmission of Dirac electrons in graphene heterojunctions}
\author{Atikur Rahman}
\author{Janice Wynn Guikema}
\author{N. M. Hassan}
\author{Nina Markovi\'c}

\affiliation{{\it Department of Physics and Astronomy, Johns Hopkins
University, Baltimore, Maryland 21218, USA.}}


%
%
%

%
\maketitle

{\bf The relativistic nature of charge carriers in graphene is
expected to lead to an angle-dependent transmission through a
potential barrier, where Klein tunneling involves annihilation of an
electron and a hole at the edges of the barrier. The signatures of
Klein tunneling have been observed in gated graphene devices, but
the angle dependence of the transmission probability has not been
directly observed. Here we show measurements of the angle-dependent
transmission through quasi-ballistic graphene heterojunctions with
straight and angled leads, in which the barrier height is controlled
by a shared gate electrode. Using a balanced differential
measurement technique, we isolate the angle-dependent contribution
to the resistance from other angle-insensitive, gate-dependent and
device-dependent effects. We find large oscillations in the
transmission as a function of the barrier height in the case of
Klein tunneling at a 45$^0$ angle, as compared to normal incidence.
Our results are consistent with the model that predicts oscillations
of the transmission probability due to interference of chiral
carriers in a ballistic barrier. The observed angle dependence is
the key element behind focusing of electrons and the realization of
a Veselago lens in graphene.}

\vskip 0.2in

{Charge carriers in graphene behave like massless,
relativistic particles \cite{gr_1, gr_3, g_2, g_3}, characterized by
chirality which arises from the existence of two interpenetrating
sublattices in the hexagonal crystal structure of graphene. Due to
the chiral nature of the charge carriers, back scattering by
impurities is forbidden \cite{back_1, back_2}, giving rise to
unusual effects in graphene p-n-p junctions such as Klein tunneling,
electron lensing and collimation \cite{klein_0, klein_1, klein_2,
klein_3, klein_4, klein_5, kk_0, kk_1, tl_1, sds_1, saj, huard,
Gordon, kim, ct_1,guiding, klein_nl, ang_1}. As a consequence of the
charge conjugation-like symmetry between electrons and holes, Klein
tunneling in graphene involves annihilation of an electron and a
hole at each p-n interface \cite{klein_3}. For a ballistic p-n-p
junction with sharp edges ($k_{F}t<1$, where $t$ is the distance
over which the potential increases at the p-n interface, and $k_F$
is the Fermi wavevector), the transmission probability is equal to
unity for normally incident charge carriers (incident angle $\phi =
0^0$, with respect to the junction normal). Away from the normal
incidence ($\phi \neq 0$), the transmission probability is expected
to oscillate as a function of the incident angle for a fixed barrier
height \cite{klein_3}. For a fixed incident angle, other than normal
incidence, the transmission probability also oscillates with varying
barrier height. However, realistic experimental configurations
typically involve smooth junctions ($k_{F}t>1 $) which focus
electrons quite effectively, and the transmission is expected to be
strongly suppressed as the incident angle increases \cite{klein_1}.
Effects of disorder \cite{disorder}and screening \cite{screening}
are also always present in realistic systems and need to be taken
into account.

\begin{figure}
    \includegraphics[width=8cm]{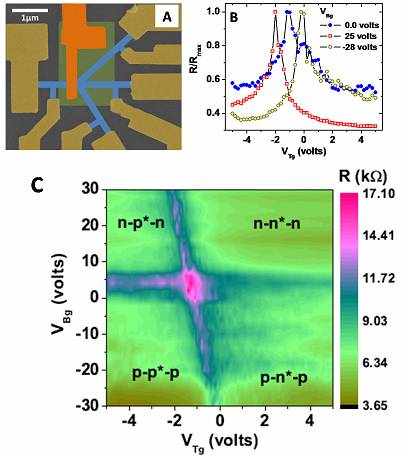}
  \caption{(A) False color SEM micrograph of a patterned
  single layer graphene device with straight and angled arms (blue). The straight arms are
perpendicular to the top gate, and the angled arms form a 45$^0$
angle with the top gate.  The
  top gate (shown in orange) is placed at the point where the straight and angled arms meet. The gate dielectric
 is shown in green (20 nm of Al$_2$O$_3$), and the leads are yellow (5nm Cr/ 75nm Au). The resistance of
each arm is measured in a four-probe configuration using non-invasive leads \cite{external_1}. The scale
bar is 1 $\mu$m long.
  (B) Normalized resistance as a function of the top gate voltage for three different values of
  back gate voltage measured on the straight arm (device SL1).
  (C) Color plot of the resistance as a function of back gate and top gate voltage
  measured on the straight arm (device SL2). The dark lines trace the shift of the charge
neutrality point as a function of $V_{Bg}$ and $V_{Tg}$,
defining regimes in which the device forms an n-p-n,
n-n-n, p-p-p or a p-n-p structure, as labeled.}
  \label{fig1}
\end{figure}

Signatures of Klein tunneling in graphene have been reported before
in gated graphene devices. Asymmetry in the resistance with respect
to the gate voltage has been attributed to the difference between
the Klein tunneling and over-barrier transmission \cite{huard,
Gordon, sonin}. Measurements on ballistic p-n-p junctions were shown
to have larger resistance than the diffusive ones \cite{klein_nl},
and electron guiding has been demonstrated based on angle-selective
transmission through a p-n interface \cite{guiding}. In the coherent
ballistic regime, conductance oscillations as a function of gate
voltage have been interpreted as tunneling through quasi-bound
states formed by Fabry-Perot interference between wavefunctions
scattered forward and backward between the two p-n interfaces
\cite{kim, ct_1, klein_4}. In contrast, direct and unambiguous
evidence of the angle dependence of transmission through a p-n-p
junction is still lacking. Sutar et al. \cite{ang_1} studied the
resistance of p-n-p junctions with angled gates, finding that the
junction resistance increases as the angle of the gate is increased
from normal incidence. However, the experiments are typically done
in the two-terminal geometry \cite{huard, Gordon, kim, ct_1,guiding,
klein_nl, ang_1}, in which case the  contact effects are unavoidable
\cite{ct_1, ct_2, ct_3, nat-nano_1,external_1} and can be difficult
to take into account. Even the four-probe measurements in graphene
can be problematic due to doping from the leads \cite{huard}. In
addition, to study the angle dependence directly, one typically has
to compare different physical samples. In such cases, direct
comparison is often obscured by inevitable differences in the
structure and the level of disorder.

Here we report a measurement of angle-dependent transmission as a
function of barrier height across graphene p-n-p (or n-p-n)
junctions. The device geometry (shown in Fig. 1A) and a balanced
differential measurement technique were specifically designed to
separate out the angle-dependent effects from other gate-dependent,
but angle-insensitive resistances.

Graphene flakes were mechanically exfoliated from natural flake
graphite and deposited on a Si wafer coated with 300 nm SiO$_2$
\cite{gr_1, gr_3}. Single-layer flakes were identified by optical
microscope and confirmed by Raman spectroscopy. Electrical leads
were patterned by standard optical and e-beam lithography and the
contacts were thermally evaporated (5nm Cr/ 75nm Au). Oxygen plasma
(100 W for 45 sec) was used to pattern the graphene. To fabricate
the top gate, 20 nm of Al$_2$O$_3$ was deposited as the top gate
insulating layer, followed by deposition of a 200 nm wide gold
electrode on top. Scanning electron microscope (SEM) image of a
typical device is shown in Fig.\ 1A. The mobility in our samples is
$\sim$ 2800 cm$^2/$Vs, which is typical for graphene on SiO$_2$
substrates. This implies an average impurity concentration of
10$^{10}$cm$^{-2}$ \cite{disorder}.

Electrical measurements were carried out at 4.2 K by placing the samples in vacuum in a He3 cryostat.
The bias current was kept sufficiently low to avoid heating. Measurements were done in a four probe
geometry with external voltage probes \cite{external_1}, using the SR 560 low noise
preamplifier and PAR 124A analog lock-in amplifier equipped with an EG\&G 116 preamplifier
(operating in differential mode).

The resistance as a function of the back gate voltage ($V_{Bg}$) and
the top gate voltage ($V_{Tg}$) is shown in Fig.\ 1B and 1C for a
typical device, measured on the straight arm at 4.2 K. Normalized
resistance as a function of $V_{Tg}$ is shown in Fig.\ 1B for three
values of $V_{Bg}$. For $V_{Bg}=0$, the resistance is symmetric
around the Dirac point, while for large positive and negative
$V_{Bg}$, it shows an asymmetry that has also been observed by
others \cite{huard}. Specifically, a larger resistance is found for
the values of $V_{Tg}$ that induce carriers of the opposite polarity
under the top gate, forming a p-n-p or an n-p-n junction. With a
suitable combination of $V_{Bg}$ and $V_{Tg}$, all portions of the
graphene device can be p or n type, or one can make a p-n-p or an
n-p-n structure, as indicated in Fig.\ 1C. The two crossing dark
lines in Fig.\ 1C trace the resistance peaks corresponding to the
Dirac points as a function of the top and back gate voltage, showing
the regions of neutrality in the sample. The slope of the diagonal
line, $dV_{Bg}/dV_{Tg}$ ($\sim 30$ in our samples), represents the
efficiency of the top gate control of the carrier density as
compared to that of the back gate.

Our graphene device has four arms, one on the left side of the top
gate (1) and three on the right side (2, 3 and 4). The top gate is
placed over arm 1, just before the point at which arms 2, 3 and 4
branch out at different angles (arm 2 at 0$^0$, and arms 3 and 4 at
$\pm$45$^0$). The resistance as a function of $V_{Tg}$ for $V_{Bg}$
= 30 volts, measured between arm 1 and each of the other three arms
(in a standard four-probe configuration using non-invasive leads
\cite{external_1}) is shown in Fig.\ 2A. Since the three current
paths share a common top-gated portion, the effect of the top gate
on all three arms is almost identical. Any angle-dependent portion
of this resistance would be included in the difference in the
resistance as a function of $V_{Tg}$ of the straight arm and either
of the angled arms, $R_{2}(V_{Tg})- R_{3}(V_{Tg})$, or
$R_{2}(V_{Tg})- R_{4}(V_{Tg})$, but \emph{not} in $R_{3}(V_{Tg})-
R_{4}(V_{Tg}$) (arms 3 and 4 are both at $\pm$45$^0$), so the
angle-dependent contribution should be identical). However, this
difference would also include other angle-independent contributions
(minor differences in size, geometry, or impurity configuration). As
is evident in Fig.\ 2A, this difference is quite small, and also
shows mesoscopic fluctuations that are typically observed in samples
of this size \cite{savchenko}.

In order to separate out the angle-dependent contribution, we use a
balanced differential measurement as shown in the schematic in Fig.\
2B. In each measurement, the current bias is applied between arm 1
and two of the arms on the right of the top gate, either one
straight and one angled arm, or two angled arms (the schematic in
Fig.\ 2B shows measurements that compare arms 2 and 3, while arm 4
is not connected). The resistances of the two paths are then
balanced with the help of a variable resistor so that the voltage
difference $\Delta$V between the two paths for $V_{Tg}=0$ is zero
(limited by the background noise of a few nanovolts). The relevant
resistances and the current branching schematic are shown in Fig.\
2C. The resistance $R_1$ of arm 1 and the resistance of the entire
p-n-p junction $R_J$ under the top gate are included in both current
paths, and the resistances $R_2=R_{2b}+R_{2d}$ and
$R_3=R_{3b}+R_{3d}$ will determine the current branching between the
arms 2 and 3 at $V_{Tg}=0$ (here $R_2b$ and $R_3b$ are the
resistances of the ballistic portions of the leads that are within
the mean free path of the right p-n interface, while $R_{2d}$ and
$R_{3d}$ are the resistances of the remaining diffusive portions of
the leads). A differential measurement of the voltage drop between
the two paths cancels out any common mode signal, including $R_1$,
$R_J$ and any external noise that affects both paths equally.
Balancing the two arms at $V_{Tg}=0$ ensures that $\Delta
V=I_2(V_{Tg}=0)R_2-I_3(V_{Tg}=0)R_3=0$, with $R_1$ and $R_J$
cancelling out (being common to both paths). Applying a voltage to
the top gate changes only the resistance under the top gate $R_J$,
which is shared by both paths and will not cause a deviation from
the balanced condition - therefore, we are \emph {not} measuring the
resistance of the p-n-p junction, or either of the two p-n
interfaces (all of which are common to both paths). A deviation from
the balanced condition, $\Delta \neq 0$, can only be caused by a
redistribution of current between arms 2 and 3 as a function of the
top gate. The current branching in arms 2 and 3 is determined by
$R_2$ and $R_3$, neither of which is directly affected by $V_{Tg}$ -
any differences in geometry, size, impurity configurations and
resistivity in the two arms are already cancelled out by balancing
$R_{2}$ and $R_{3}$, and neither arm is under the top gate. If,
however, there were any gate-dependent difference in the
transmission probability between the two arms, it would affect the
distribution of current in the arms 2 and 3 within the distance from
the junction of the order of the mean free path. More specifically,
it would affect the current distribution through $R_{2b}$ and
$R_{3b}$, causing a difference in the voltage drop over $R_{2b}$ and
$R_{3b}$. This would cause a deviation $\Delta V(V_{Tg})$ from the
balanced condition, which measures the difference in the voltage
drop \emph{only} over the ballistic portion ($\Delta V=
I_2(V_{Tg})R_{2b}-I_3(V_{Tg})R_{3b}$). It is this deviation from the
balanced condition that we measure below. In order to relate the
measurement to the transmission probability, we define a parameter
$R_{dev}=\Delta V(V_{Tg}=0)/I_1$ with units of resistance. $R_{dev}$
is then proportional to the difference in the transmission
probability and reflects the current redistribution in the two leads
as a function of $V_{Tg}$.

\begin{figure}
     \includegraphics[width=8cm]{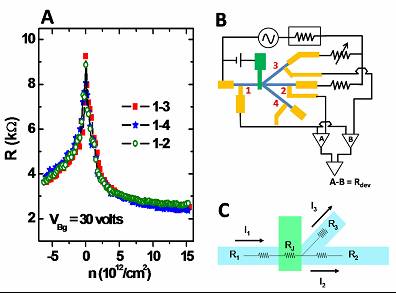}
    \caption{(A) Resistance as a function of carrier density under the application of top gate voltage for arms 2, 3
    and 4 on device SL1. The resistance on each arm was measured using standard
    four-probe technique with non-invasive leads. Nearly identical resistance
    curves as a function of $V_{Tg}$ on all three arms reflects the fact that
    all three arms share the same gated portion. In order to reliably measure
    the difference between pairs of curves shown here, we use the balanced
    differential measurement technique shown in (B).
(B) Measurement schematic used to isolate the angle-dependent
portion of the resistance. In this schematic, the current branching
in the arms 2 and 3 is balanced (using a variable resistor) in such
a way that the voltage difference between A and B is zero when
$V_{Tg}$=0 (arm 4 is not connected on this schematic). The deviation
from the balanced condition, $\Delta$ V is then measured as a
function of $V_{Tg}$. In order to relate this measurement to the
transmission probability, we define a parameter $R_{dev}=\Delta
V(V_{Tg}=0)/I_1$ with units of resistance. $R_{dev}$ is then
proportional to the difference in the transmission probability and
reflects the current redistribution in the two leads as a function
of $V_{Tg}$. (C) Current branching diagram for arms 2 and 3. $R_1$
and $R_J$ denote resistances of arm 1 and the entire junction
defined by the top gate, respectively. $R_1$ and $R_J$ are both
common to both current paths. The resistances of arms 2 and 3 can be
further separated into ballistic and diffusive parts, so that
$R_2=R_{2b}+R_{2d}$ and $R_3=R_{3b}+R_{3d}$. Here $R_{2b}$ and
$R_{3b}$ denote the resistances of the portions of arms 2 and 3 that
are within the mean free path of the junction, while $R_{2d}$ and
$R_{3d}$ are the resistances of the rest of the arms 2 and 3.}
  \label{fig2}
\end{figure}

For a particular back gate voltage, we begin by balancing two of the
arms (2 and 3) at $V_{Tg}$=0. Then, we study $R_{dev}$ as a function
of $V_{Tg}$, shown in  Fig.\ 3A (the balancing was done at $V_{Bg}$
= -28 volts and $V_{Tg}$ =0). $R_{dev}$ shows reproducible
fluctuations throughout the whole range of $V_{Tg}$, which resemble
mesoscopic conductance fluctuations \cite{savchenko} (the amplitude
of the fluctuations in units of conductance are on the order of 0.01
$e^2/h$). The amplitude of the fluctuations in $R_{dev}$ visibly
increases for positive values of $V_{Tg}$, when we expect a p-n-p
junction to have formed. If we balance the circuit at $V_{Bg}$ = 25
volts and $V_{Tg}$ =0, larger fluctuations in  $R_{dev}$ are
observed below $V_{Tg}$ = -2 volts, when we have an n-p-n junction
formed under the top gate (see Fig.\ 3B). In order to emphasize the
amplitude of these fluctuations, we examine the derivative of
$R_{dev}$ as a function of $V_{Tg}$ ($dR_{dev}/dV_{Tg}$), as shown
in Fig.\ 3C. $dR_{dev}/dV_{Tg}$ is shown here as a function of
$V_{Tg}$, along with the corresponding change in resistance of the
entire path to illustrate the proximity to the Dirac point. It is
evident that the fluctuations are significantly larger in the p-n-p
side than in the p-p-p side. Similarly, larger fluctuations are
observed in the n-p-n side compared to the n-n-n side of the top
gate voltage axis (Fig.\ 3D).

\begin{figure}
\includegraphics[width=8cm]{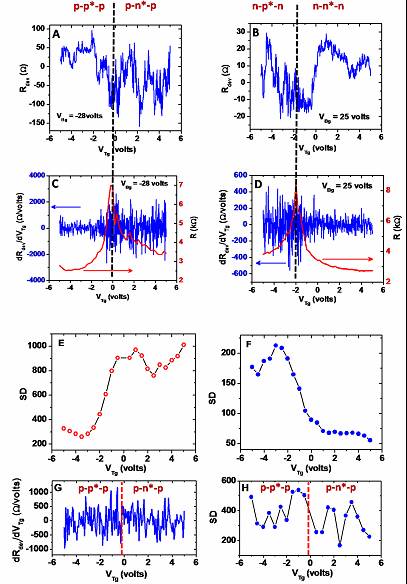}
    \caption{$R_{dev}$ as a function of $V_{Tg}$ for (A) $V_{Bg}$=-28 volts and (B) $V_{Bg}$=25
    volts. $R_{dev}$ is measured by balancing arms 2 and 3, as shown in the schematic in
Fig.\ 2(B). Corresponding derivatives of $R_{dev}$ with respect
to $V_{Tg}$ are shown in (C) for $V_{Bg}$=-28 volts and (D) for
$V_{Bg}$=25
    volts (left axis). The corresponding change in resistance is shown in red (right
    axis). The amplitude of the fluctuations is visibly larger
in the p-n-p and n-p-n regimes than in the p-p-p and n-n-n regimes.
Standard deviation of $dR_{balance}/dV_{Tg}$ as a function of $V_{Tg}$ is shown in
  (E) $V_{Bg}$ = - 28 volts and (F) $V_{Bg}$ = 25 volts. The standard deviation of
   $dR_{balance}/dV_{Tg}$ increases up to four times as the variation of $V_{Tg}$ takes
the device from the p-p$^*$-p to p-n$^*$-p or from n-n$^*$-n to
n-p$^*$-n regime. (G) A derivative of $R_{dev}$ with respect to
$V_{Tg}$ and its standard deviation (H) measured by balancing arms 3
and 4 (both at at $\pm$45$^0$ with respect to arm 2) at $V_{Bg}=
-28$ volts. No change is observed as the device enters the bipolar
regime in this case. }
  \label{fig3}
\end{figure}

To obtain a more quantitative measure of the amplitude of these
fluctuations, we divided the data points into small bins and
calculated the standard deviation of $dR_{dev}/V_{Tg}$ as a function
of $V_{Tg}$. Figures 3E and 3F show the standard deviation
calculated with twenty five data points per bin and averaged over
four data points. It is evident that the standard deviation of the
fluctuations increases by a factor of four as we cross from p-p-p to
p-n-p or from n-n-n to n-p-n region (we note that the standard
deviation of the fluctuations is overall larger when we have a p-n-p
junction, than in the case of a n-p-n junction).

Similar results were obtained by comparing arms 2 and 4 (one straight and one angled arm). However, if we
compare arms 3 and 4 (both at 45$^0$ angle
with respect to the junction interface, we find that the variation in the
$R_{dev}$ remains nearly constant throughout the whole range of top and back gate voltages, as shown in Fig. 3G.
This is also confirmed by the standard deviation shown in Fig. 3H.

\begin{figure}
    \includegraphics[width=8cm]{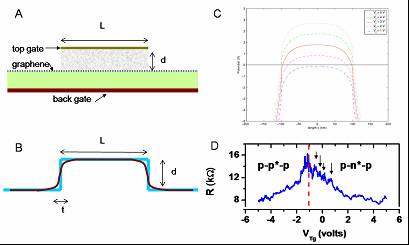}
  \caption{(A) Side view schematic of the top gated portion of the sample.
  (B) Shape of the potential barrier formed by
the top gate. (C) Potential profiles under the top gate calculated
by solving the Laplace equation with appropriate boundary
conditions. (D) Resistance as a function of $V_{Tg}$ measured on the
straight arm 2. The arrows point at the local peaks in the
resistance. }
  \label{fig4}
\end{figure}

There are two possible contributions to $R_{dev}$ that could depend
on $V_{Tg}$: mesoscopic conductance fluctuations and angle-dependent
transmission. The balanced differential measurement cancels out any
other contributions due to the size, geometry, and other
contributions that do not depend on the gate voltage. Mesoscopic
conductance fluctuations are indeed expected in samples of similar
size \cite{savchenko}. These fluctuations are the consequence of
quantum interference of electron wavefunctions scattered on
impurities, and are reproducible as a function of gate voltage for a
given impurity configuration. The top gate voltage in our sample
does not affect the bulk of arms 2, 3 and 4, but it is likely to
affect a small portion within the mean free path of the top gate.
Since the top gate has a finite width, the impurity configuration
along this width, while certainly similar, is not exactly identical.
Therefore, $dR_{dev}$ would measure the difference in conductance
fluctuations in small ballistic portions of two different arms. This
would manifest as random, but reproducible fluctuations in
$dR_{dev}$ as a function of $V_{Tg}$, as is indeed observed in our
measurement in the entire range of $V_{Tg}$. In the case when a
bipolar junction is formed under the top gate, the fluctuations are
visibly increased. To see whether this increase reflects enhanced
mesoscopic conductance fluctuations, we examine the $dR_{dev}$
measured by balancing different pairs of arms. Measuring all pairs
of arms on three different devices, we find that increased
fluctuations are observed only when the two arms are at a different
angle - if we compare arms 3 and 4, the amplitude of the
fluctuations remains the same throughout the whole range of
$V_{Tg}$. We therefore conclude that the difference in the
mesoscopic conductance fluctuations in the ballistic portions of the
leads is the cause of the baseline fluctuations, but cannot account
for the increase in the fluctuation amplitude (leading to the
four-fold increase in the standard deviation) observed when a
bipolar junction has formed under the top gate.

In order to examine the possibility that this increase is due to
angle-dependent difference in the transmission detected on different
arms of our devices, we need to consider the appropriate model for
the potential barrier formed by the top gate (Fig 4A). A sharp,
ballistic, rectangular barrier of length L and height d (shown in
Fig. 4B in blue) would lead to perfect transmission at normal
incidence and an oscillating transmission probability at other
incident angles. A differential measurement of one straight and one
angled arm would then result in oscillating resistance in the case
of a bipolar junction, and no oscillations in the case of
over-barrier transmission (p-p-p or n-n-n case). These oscillations
would be superposed with the mesoscopic conductance fluctuations,
resulting in increased fluctuation amplitude in the case of bipolar
junction. This could, in principle, explain our results. However,
given that the estimated mean free path is on the order of the top
gate length, our devices are quasi-ballistic at best and it is not
obvious that we can assume the potential to be sharp. Estimating the
mean free path ($l_{mfp}$) from its relation to conductance
$\sigma=2 (e^2/h)(k_{F}l_{mfp})$, we find it to be around $\sim$
100-200nm in our samples ($l_{mfp}=110nm$ for the sample shown in
Fig 2 A, evaluated at 3V). The width of the top gate in all samples
is 200 nm. Therefore, we cannot assume that the entire p-n-p
structure is ballistic, but we can think of it as quasi-ballistic,
as some portion of carriers is likely to pass through the barrier
without scattering. Alternatively, we may have to treat the p-n-p
junction as two p-n junctions in series, as discussed below. A
realistic barrier, however, would not have infinitely sharp edges -
we expect the potential to rise over some distance $t$ (Fig. 4B).
The edges can still be considered sharp if the electron wavelength
is large compared to $t$, or if $k_{F}t<1$. In our samples, $k_{F}t
\sim 10$ (taking $k_{F}= \sqrt{\pi n}$ at 3V, away from the Dirac
point), so we need to consider smooth edges.

Another important consideration is whether transport through each
p-n junction is ballistic or diffusive. A good measure for this is a
parameter $\beta=n'/n_{i}^{2/3}$, where n' is the density gradient
at zero energy, and $n_{i}$ is a parameter related to mobility
through $n_{i}=e/h\mu$ \cite{disorder}. In order to observe
ballistic transport through a p-n junction, it is required that
$\beta>10$. The mobility can be estimated from $\mu=1/en\rho(n)$,
where n is the carrier density and $\rho(n)$ is the resistivity of
the sample away from the Dirac point. The mobility in our samples is
$\sim$ 2800 cm$^2/$Vs, and the top gate is placed at the height
d=20nm away from graphene, as shown in Fig. 4B. The distance over
which the potential increases at the p-n interface, t, can be taken
to be on the order of d. Estimating the density gradient at the
interface from the difference in the carrier densities on the two
sides of each p-n junction over 20nm, we arrive at the value of
$\beta\sim$160, indicating that we should expect a significant
ballistic contribution. This is confirmed by the calculation of the
potential profiles under the top gate, shown in Fig. 4C.

A model for smooth ballistic p-n junctions was developed in
reference  \cite{klein_1}. This model predicts that a smooth p-n
junction transmits only the carriers that approach it within a small
incident angle $\theta$ smaller than $\theta_{0}=(\pi k_{F}t)$. The
transmission probability as a function of the incident angle is
\cite{klein_1}: $T(\theta)=e^\pi(k_{F}t \sin^2 \theta)$ so we would
expect perfect transmission for normally incident carriers, and
suppressed transmission with increasing incident angle. In our
samples $\theta_{0}= 6^0$, so we would expect efficient collimation
within the mean free path away from the p-n interface. Our p-n-p (or
n-p-n) junctions include two p-n interfaces - one on either edge of
the top gate. The first p-n junction would then preferentially
transmit more carriers that are nearly normally incident to the
barrier. These carriers would arrive to the second p-n junction,
which would again select more of the normally incident ones. As all
the arms have the same width, and the branching point is less than
$l_{mfp}$ away from the top-gated portion, most of the electrons
coming from arm 1, that emerged in the direction normal to the
second pn junction, will go straight into arm 2. The arm 3 (or 4) is
placed at an angle with respect to the top gate ($\phi = 45^0$), so
it will preferentially collect any electrons that emerged at angles
close to $\phi = 45^0$. Here we also have to take into account the
fact that the arms have a finite width, and the electrons can emerge
anywhere along that width. Therefore, each arm will actually collect
electrons that emerge at a small range of angles.

If we assume that the carriers tunnel through two independent p-n
junctions in series, than we would expect most of the carriers to be
collected by arm 2, with very few carriers going into arm 3 or 4. In
this case, we would not expect to see oscillations in the
transmission amplitude as a function of barrier height.

If, however, the transport through the p-n-p junction is coherent,
one can expect to observe resonant tunneling through quasi-bound
states due to Fabry-Perot interference \cite{klein_4, kim}. In this
case, the oscillating part of the resistance can be approximated by
\cite{klein_4, kim}: 
\begin{equation} 
R_{osc} ^{-1}=\frac{8e^2}{h}\sum 
[T^4 (1-T^2) \cos \theta_{WKB} e^{-\frac{2L}{l_{mfp}}}]\end{equation}
where T is the transmission coefficient and  $\theta_{WKB}$ is the semiclassical phase \cite{klein_4, kim}.
Taking into account nonlinear screening when evaluating T
\cite{screening}, this expression results in resistance oscillations
with an approximate period in the carrier density of about $10^{12}$
cm$^{-2}$. In our samples, this would result in oscillations with
$V_{Tg}$ with a period of about 0.5V, which is indeed observed in
four-terminal measurements of resistance on arm 2, as shown in Fig 4D. 
These oscillations are superimposed with the mesoscopic
conductance fluctuations, as expected from Fabry-Perot resonances in
the presence of disorder \cite{sds_1}. In the presence of disorder,
it was found that the oscillations survive even at impurity
concentrations that are by an order of magnitude larger than that
observed in our samples, but become smeared by mesoscopic
conductance fluctuations. Such fluctuations, in addition to
inhomogeneous gate coupling due to the disorder, were also found to
cause averaging over several Fabry-Perot fringes in all but the
cleanest samples \cite{kim}. In this picture, small-angle averaged
irregular oscillations would be expected on arm 2 when a bipolar
junction is formed under the top gate, with very few carriers making
it to arm 3 or 4. Comparing arm 2 with either 3 or 4 would then
result in the increased oscillation amplitude, while comparing arms
3 and 4 would not, as observed in our experiment.

We note that it is not necessary for the entire device to be
ballistic in order to observe the angle dependence of the
resistance.  As long as the difference in the current distribution
due to ballistic effects in a small portion of the straight and
angled arms is large enough, it will manifest as the deviation from
the balanced condition with changing top gate voltage. Therefore,
our measurement technique specifically \emph{selects} and only
measures the ballistic contribution. Since all the arms share the
same top-gated portion, the properties of the potential barrier are
identical for straight and angled arms: there will be no differences
in the pn-junction length, roughness, the nature of disorder,
contact resistance and other issues that have to be taken into
account when comparing different physical devices.

In conclusion, even though our measurements were limited by mesoscopic conductance fluctuations,
we were able to observe angle-dependent transmission through a p-n-p junction in graphene.
Using cleaner samples with higher mobility and a larger mean free path would make it possible to make
more detailed angle-resolved measurements, which could lead to electron optics applications.

\end{document}